\documentclass[10pt]{article}
\usepackage{opex3}
\begin{document}
\title{
Moment-generating function method used to accurately evaluate the 
\\
impact of the linearized optical noise 
amplified by EDFAs  
}
\author{Zhongxi Zhang$^*$, Liang Chen, Xiaoyi Bao}
\address{Department of Physics, University of Ottawa, 150 Louis Pasteur, Ottawa, K1N 6N5,  Canada \\
        }
\email{$^*$zhongxiz@yahoo.com}

\begin{abstract}
In a nonlinear optical fiber communication (OFC) system with signal power much stronger  than noise power,
the noise field in the fiber can be described by linearized noise equation (LNE). In this case,
the noise impact on the system performance can be evaluated by  moment-generating function (MGF) method.
Many published MGF calculations were based on the LNE using
continuous  wave (CW) approximation, where the modulated signal needs to be
artificially simplified as an unmodulated  signal.
Results thus obtained should be treated carefully.  
More reliable results can be obtained by improving the CW-based LNE with
the accurate LNE proposed by
Holzl$\ddot{\mbox{o}}$hner {\it {et al}} in Ref. \cite{Holzlohner02}.
In this work we show that, for the case of linearized noise amplified by EDFAs, its MGF can be
calculated by obtaining the noise propagation information directly from the accurate LNE.
Our results agree well with the experimental data of multi-span DPSK systems.
%
%
%
\end{abstract}
\vskip 0.10 in

\section{Introduction}
\label{Introduction}

The amplified spontaneous emission (ASE) noise from optical amplifiers,
e.g.,
Erbium Doped Fiber Amplifiers (EDFAs), is one of the fundamental reasons for the
bit-error-rate (BER) in an optical fiber communication (OFC) system.
For  an OFC system with non-negligible Kerr nonlinearity,
the ASE impact evaluation is complicated due to the
nonlinear interaction between signal and ASE noise.
In the case of signal power much stronger than the noise power, the noise-noise  beating is relatively small so that
the noise field in the fiber can be approximately described by
linearized noise equation (LNE), which was proposed separately 
in Ref. \cite{Holzlohner02} and Refs. \cite{Kikuchi93}-\cite{Secondini09}.

Noise propagator is a matrix used to show noise field propagation in the fiber.
In the case of linear perturbation, it is independent of the specific noise realizations, which makes it possible
to calculate the moment-generating function (MGF) of the filtered photoelectric current
at  the receiver \cite{Holzlohner02}-\cite{Secondini09}.

MGF method is an approach making use of MGF 
to evaluate the noise impact on the system performance.
Now, it is  well known that this method is accurate for various linear OFC systems.
For nonlinear OFC systems, the MGF method can also be
applied, provided  their noise fields obey LNE.
One can see this from  Doob's theorem, which means that, in a linearizable  system driven by Gaussian-distributed noise, each of the independent random variable keeps Gaussian (cf. P. 35 of Ref. \cite{Holzlohner_phd}).
Thus the MGF of the received photoelectric current can be calculated as those for  linear OFC systems.

The common form of LNE \cite{Kikuchi93}-\cite{Secondini09} is based on the continuous  wave (CW) assumption, i.e.,
the noise-free  signal in the LNE is artificially simplified as a CW wave. 
As a result, a semi-analytical form of noise propagator and the noise power spectral density (PSD),
so  called parametric gain (PG), can be obtained.
The drawback 
of this simplification is that the noise-free signal in this LNE neglects chromatic dispersion (CD) effect.
As a result,  the couplings between noise components (in frequency domain) 
cannot be taken into account. 

A LNE beyond CW, named accurate LNE in this work, was first proposed and discussed in Ref. \cite{Holzlohner02}.
Dynamically taking into account the local CD and Kerr nonlinearity along the fiber,
this LNE provides accurate noise information,  
with its computational cost being
much higher than the CW approach.
For example, given  a noise-free signal obtained from nonlinear Schr$\ddot{\mbox{o}}$dinger equation (NLSE), 
the computation required to update the accurate LNE has cubic complexity in the number of Fourier components \cite{Holzlohner02}.
To  reduce the computational complexity,  
covariance matrix method (CMM) was proposed in Ref. \cite{Holzlohner02}, where the noise covariance matrix
was obtained by processing large  noise realizations.
In Refs. \cite{Holzlohner02b,Holzlohner03}, the computational cost of CMM was further reduced by
a deterministic approach using perturbation solution.
Since the raw covariance matrix obtained from NLSE via
Monte Carlo noise realizations \cite{Holzlohner02} or perturbation solution \cite{Holzlohner02b,Holzlohner03}
may contain
nonlinear noise contribution,
it needs to be separated from
the nonlinearity-induced  phase and timing jitter.
Thus, the obtained
pdfs of the receiver voltage
agrees well with Monte Carlo simulation \cite{Holzlohner02,Holzlohner02b,Holzlohner03b}.

With the help of linear perturbation, the noise covariance matrix 
can also be solved from  its ordinary differential equation (ODE) proposed by
\cite{Holzlohner02, Demir04}. 
The covariance matrix obtained by solving such linear ODE does not need  jitter separation, although this
ODE is more complicated than the accurate LNE \cite{Demir04}.
So far there is little comparison between the approaches of Ref. \cite{Demir04} and Refs. \cite{Holzlohner02, Holzlohner02b,Holzlohner03}.

In this work,  we simplify the CMM by showing that the noise propagator matrix can be obtained directly from the accurate LNE.
Therefore, there is no nonlinearity-induced jitter.
To effectively reduce the computational complexity in updating the LNE, one can
decompose the Kerr effect related matrix into a symmetric and an antisymmetric matrices
[cf. the discussion after Eq. (\ref{hat_L_nu_mu}) in Appendix A].
Making use of the fourth-order Runge-Kutta in the interaction picture
(RK4IP) method \cite{Hult07,zcb10}, the accurate LNE can be solved with large step size, as detailed in Sec. \ref{Noise_propagator}.
We evaluate the impacts of noise propagator
on moment-generating function (MGF) and BER
in  Sec. \ref{MGF_1}.
The accuracy of this new approach depends on how far the linearized noise 
deviates from the actual noise.
To numerically verify this new approach, we consider the BERs in  a 20-span DPSK system with
nonlinear phase of $\bar{\Phi}_N=0.2\pi$ \cite{Serena06} in Sec. \ref{N_verification}.
Our BER calculations agree well with the published  CMM results. 
In Sec.  \ref{N_verification},
we also simulate the experiments of the multi-span DPSK systems discussed in Ref. \cite{Coelho09}  and show that,
to fit the experimental data, one needs to take into account the
nonlinearity induced phase difference between noise and noise-free signal, which  will affect the signal-noise beating significantly.

\section{Noise propagator obtained from the accurate LNE}
\label{Noise_propagator}
In an OFC system amplified by EDFAs, the noise propagator is a fundamental matrix that 
determines the noise impacts on MGF and  BER.
In this section, we show that 
the noise propagator matrix in a fiber of length $L$ can be obtained from the accurate LNE given by Eq. (\ref{NS_EDFA}).
For a multi-span OFC system,  one needs to introduce an equivalent noise propagator which can be obtained from  
PG. 

\subsection{Noise propagator in a fiber of length $L$}

The noise propagator in a fiber of length $L$ can be obtained by extending the
RK4IP in Refs. \cite{Hult07,zcb10} to the accurate LNE given by Eq.  (\ref{NS_EDFA}) in Appendix A, 
where the linear operator $\hat{L}$ is associated with CD effect, 
whereas the nonlinear operator $\hat{N}$ is caused by Kerr nonlinearity.
By introducing   
$\tilde{a}=e^{\hat{L}(z-z_0)}\tilde{a}^I$ and
$\hat {N}^I=e^{-\hat{L}(z-z_0)}\hat{N}e^{\hat{L}(z-z_0)}$,
Eq. (\ref{NS_EDFA}) in the interaction picture (IP) has the form
\vskip  -4mm
\begin{eqnarray}
\frac{d \tilde{a}^I}{d z}
=\hat{N}^I \tilde{a}^I
\label{NS_EDFA_IP}
\end{eqnarray}
 \vskip  -1mm
Taking $z_0=z_n+h/2$ with step size $h=z_{n+1}-z_{n}$ and denoting $\tilde{a}_n=\tilde{a}(z_n)$, $\tilde{a}_{n+1}=\tilde{a}(z_{n+1})$,
$\hat{N}^I_n=e^{\hat{N}h/2}\hat{N}(z_n)e^{-\hat{N}h/2}$, $\hat{N}^I_{n+1/2}=\hat{N}(z_n+h/2)$, and
$\hat{N}^I_{n+1}=e^{-\hat{L}h/2}\hat{N}(z_{n+1})e^{\hat{L}h/2}$, one can
use RK4IP \cite{Hult07,zcb10} to
to solve Eq. (\ref{NS_EDFA_IP}) with
\vskip  -5mm
\begin{eqnarray}
&&\tilde{a}_{n+1}=e^{\hat{L}h/2}[\tilde{a}_n^I+\frac{hk_1}{6}+\frac{hk_2}{3}+\frac{hk_3}{3}+\frac{hk_4}{6}]
\nonumber\\
&&\tilde{a}_n^I=e^{\hat{L}h/2}\tilde{a}_n
\nonumber\\
&&k_1=\hat{N}_n^I\tilde{a}_n^I=e^{\hat{L}h/2}\hat{N}(z_n)\tilde{a}_n\equiv \hat{k}_1\tilde{a}_n
\nonumber \\
&&k_2=\hat{N}^I_{n+1/2}[\tilde{a}_n^I+\frac{hk_1}{2}]=\hat{N}(z_n+h/2)[e^{\hat{L}h/2}+\frac{h\hat{k}_1}{2}]\tilde{a}_n\equiv \hat{k}_2\tilde{a}_n
\nonumber \\
&&k_3=\hat{N}^I_{n+1/2}[\tilde{a}_n^I+\frac{hk_2}{2}]=\hat{N}(z_n+h/2)[e^{\hat{L}h/2}+\frac{h\hat{k}_2}{2}]\tilde{a}_n\equiv \hat{k}_3\tilde{a}_n
\nonumber \\
&&k_4=\hat{N}_{n+1}^I[\tilde{a}_n^I+hk_3]=e^{-\hat{L}h/2}\hat{N}(z_{n+1})e^{\hat{L}h/2}[e^{\hat{L}h/2}+h\hat{k}_3]\tilde{a}_n
\equiv \hat{k}_4\tilde{a}_n
\label{H_RK4IP}
\end{eqnarray}
\vskip -3mm
or
\vskip -3mm
\begin{equation}
\tilde{a}_{n+1}=\Bigg(e^{ \hat{L}h/2}\bigg[e^{\hat{L}h/2}
                                         +\frac{h\hat{k}_1}{6}
                                         +\frac{h\hat{k}_2}{3}
                                         +\frac{h\hat{k}_3}{3}\bigg]
             +\frac{h}{6}\hat{N}(z_{n+1})e^{\hat{L}h/2}(e^{\hat{L}h/2}+h\hat{k}_3)\Bigg)\tilde{a}_n,
\label{H_RK4IP_2}
\end{equation}
which means
the noise propagator for the fiber of length $h=z_{n+1}-z_{n}$ can be calculated as
\begin{equation}
H(z_{n+1},z_n)=e^{\hat{L}h/2}\bigg[e^{\hat{L}h/2}
                           +\frac{1}{3}(\frac{h}{2}\hat{k}_1+\frac{2h}{2}\hat{k}_2+h\hat{k}_3)\bigg]
            +\frac{h}{6}\hat{N}(z_{n+1})e^{\hat{L}h/2}\bigg[e^{\hat{L}h/2}+h\hat{k}_3\bigg]
\label{H_h}
\end{equation}
For the fiber of length $L$, the  noise propagator has the form
\begin{equation}
p_n(L,0)=H(L,L-h_L)\cdots H(h_1,0).
\label{p_noise_L}
\end{equation}

Note that the RK4IP used here is different from the RK4IP in Ref. \cite{zcb10}, 
where what to be solved  was the noise-free signal (a 1D matrix), 
while here what we want is the noise propagator (a 2D matrix).
The computational complexity for this 2D matrix is $O(N_w^3)$,
due to that each $\hat{k}_i$ ($i=2,3,4$) in Eq. (\ref{H_RK4IP}) 
needs one dense matrix multiplication.
Here $N_w$ is the number of Fourier components used for signal representation, as mentioned between Eqs. (\ref{hat_L_nu_mu}) and (\ref{auto_noise_f_W}).

\subsection{Equivalent noise propagator of a multi-span system}
As discussed in Appendix A, the ASE from an EDFA can be modeled as additive white Gaussian noise (AWGN) 
with its variance given by Eq. (\ref{auto_noise_f3}).
Given the ASE injected at the input of a fiber and the noise propagator  
obtained from Eqs. (\ref{H_RK4IP}), (\ref{H_h}), and  (\ref{p_noise_L}), the noise PSD (or PG) 
at the output of a fiber of length $L$ can be written as \cite{Holzlohner02,Serena06,Coelho09,Secondini09}
\begin{equation}
PG_1=p_n(L,0)\sigma^2 I p_n^{T}(L,0)=\sigma^2 p_n(L,0)p_n^{T}(L,0),  
\label{PG_1}
\end{equation}
where $I$ is a unit matrix and Eq. (\ref {auto_noise_f3}) has been used. In Eq. (\ref{PG_1}), $p_n^{T}$ is the transpose of $p_n$. 
 
For a $K$-span system consisting of $(K+1)$ EDFAs (cf. Fig. \ref{fig_link}), its PG has the form 
\begin{eqnarray}
&&
\widehat{PG}=\big( G\sigma_{in}^2+\sigma^2\big) P_n(K,0)P_n^T(K,0) 
+ \sigma^2 \sum_{k=1}^{K} P_n\big{(}K,k\big{)}P_n^T\big{(}K,k\big{)}, 
\nonumber\\
&&
P_n\big{(}K,k\big{)}=p_n\Big{(}KL,(K-1)L\Big{)}\cdots p_n\Big{(}(k+1)L,kL\Big{)},\;\; \;\;\;(k=0,1,\cdots,K-1)
\nonumber\\
&&
P_n\big{(}K,K\big{)}=I,
\label{PG_K}
\end{eqnarray}
where 
$\sigma_{in}^2=N_{\mathrm{in}}/(2T_0)$ and $\sigma^2$ is given by Eqs. (\ref{PSD_ASE}) and (\ref{auto_noise_f3}). 
In Eq. (\ref{PG_K}), the real symmetric matrix $\widehat {PG}$ is positive definite. It can be factorized as
\begin{equation}
\widehat {PG}=\sigma^2 P_{n,eq} P_{n,eq}^{T}, 
\label{PG_K_eq}
\end{equation}
where the equivalent noise propagator $P_{n,eq}$ can be obtained either by using Cholesky decomposition or 
symmetric (square root) decomposition \cite{Mathai92}. 
The latter yields $P_{n,eq}=P_{n,eq}^{T}$.
\section{MGF calculation}
\label{MGF_1}
With the noise propagator matrix, one can  evaluate the BER in the OFC system by calculating
the MGF of the electrically filtered current $I(t_{\mathrm{s}})$
expressed using Karhunen-Lo$\grave{\mbox{e}}$ve series expansion (KLSE).



Due to the noise in the OFC system, the received (or filtered photoelectric) current fluctuates around its expectation value.
The MGF of such current is a useful form of its  probability distribution.
To get a simple form of MGF, the received  current needs to be expressed using KLSE.  
For a nonlinear OFC system with its noise being linearizable, the KLSE form of the received current can be formulated as Eq. (\ref{y_receiv2}) 
and the related formulas in Appendix B.
All the parameters for the nonlinear case are generalized from those for the linear case, which 
has been well discussed in Refs. \cite{Forestieri00,zcb07b} and other
publications. In Eqs. (\ref{tilde_s_n}) and (\ref{y_receiv2}), the Dirac notation $|\tilde{Z} \rangle$ is used to represent  the normalized noise 
(in Re-Im form) expressed by Karhunen-Lo$\grave{\mbox{e}}$ve bases (in Re-Im form).   
Averaging $|\tilde{Z}\rangle_i$ 
($i=1,\cdots,4M_n+2$) with  
formula \cite{Forestieri00,zcb07b} 
\begin{eqnarray}
E\Big[ \exp\Big( s( \tilde{\lambda} c^2+2c\tilde{b})  \Big) \Big]=\!\!
\int_{-\infty}^{\infty}\!\!\frac{dc}{\sqrt{2\pi\sigma^2}}\exp(-\frac{ c^2}{2\sigma^2})\exp [ s(\tilde{\lambda} c^2+2c\tilde{b}) ]
=\frac{ \exp[\frac{2\sigma^2 s^2 \tilde{b}^2}{1-2\sigma^2 s \tilde{\lambda}} ] } {\sqrt{1-2\sigma^2s\tilde{\lambda }}},
\label{gaus_int}
\end{eqnarray}
the MGF of the filtered current, denoted as $\Psi_{t_{\mathrm{s}}}(s)$ here, can be written as
\begin{eqnarray}
\Psi_{t_{\mathrm{s}}}(s)=E\Big[ \exp [sI(t_{\mathrm{s}}) ] \Big]
= \exp[  sI_{\mathrm{ss}}(t_{\mathrm{s}}) ] \prod_{i=1}^{4M_n+2}\frac{\exp[\frac{2\sigma^2 s^2 \tilde{b}_i^2(t_{\mathrm{s}})}{1-s\beta_i}]}
{(1-s\beta_i)^{\xi}},\;\;\;\;(\beta_i=\!2\sigma^2\tilde{\lambda}_i)
\label{MGF}
\end{eqnarray}
where 
$I(t_{\mathrm{s}})$ is the filtered photoelectric current at time $t_{\mathrm{s}}$. It consists of
signal-signal beating ($y_{ss}$), noise-noise beating ($y_{nn}$), and signal-noise beating ($y_{ns}$), which are
detailed
in  Eqs. (\ref{y_D}) and (\ref {y_receiv2}) respectively. In Eq. (\ref{MGF}),
$\tilde{b}_i(t_{\mathrm{s}})$ is the $i$th component of $|\tilde{b}(t_{\mathrm{s}})\rangle$ in  (\ref {y_receiv2}),
while  $\tilde{\lambda}_i$ is the power of $i$th component of the noise in Karhunen-Lo$\grave{\mbox{e}}$ve presentation.
In Eq. (\ref{MGF}), $\sigma^2$ is given by Eq. (\ref{auto_noise_f3}).  
In this work, we take $\xi=1/2$ 
for polarized noise.  

With the help of Eq. (\ref {MGF}) as well as Eqs. (7) and (8) in Ref. \cite{zcb07b}, the BER can be calculated.  


\section{OSNR at the receiver}
\label{RX_OSNR}

For an optical system with ASE power being much larger than other noise sources, the OSNR
with reference bandwidth $B_r$ (0.1nm) 
can be calculated as
\begin{equation}
OSNR_{0.1nm}=\frac{\bar{P_s}}{P_{ASE}(B_r)}.
\label{OSNR}
\end{equation}
In Eq. (\ref{OSNR}), $\bar{P_s}$ is the time-averaged (noise free) signal power,
while $P_{ASE}(B_r)$ is the noise power within $B_r$.
To obtain  $\bar{P_s}$ and $P_{ASE}(B_r)$, one needs to notice that
the measurement bandwidth $B_m$ 
[e.g., the bandwidth of the transfer function of an optical spectrum analyzer (OSA)] 
may  not be the same as $B_r$. 
Thus the $\bar{P_s}$ in Eq. (\ref{OSNR}) becomes the power of the signal filtered by $B_m$,
while $P_{ASE}(B_r)$ becomes the ASE filtered by $B_m$ and weighted by a factor $B_r/B_m$ \cite{Gariepy09}.  

In a linear optical system, 
the ASE noise along the fiber can be treated as AWGN. Thus, for the system of Fig. \ref{fig_link}, 
its OSNR can be simply calculated as 
\begin{equation}
OSNR_{L,0.1nm}=\frac{\bar{P_s}}{P_{ASE}(B_m)}\times\frac{B_m}{B_r} 
\;\;\;\;\;\; \Big{(}P_{ASE}(B_m)=[GN_{in}+N_0(K+1)]B_m\Big{)},
\label{OSNR_L}
\end{equation}
where $P_{ASE}(B_m)$ is the ASE power within $B_m$ and $N_0$ is given by Eq. (\ref{PSD_ASE}). 
In Eq.(\ref{OSNR_L}), the filter ($B_m$) effect on the ASE has been neglected.

In a nonlinear optical system, the ASE noise ``amplified by'' PG cannot be treated as a white noise.  
Similar to the noise-noise beating given in Eq. (\ref{y_receiv2}), 
the measured  ASE power will only relate with the self-beating terms of the noise, which yields
\begin{equation}
OSNR_{NL,0.1nm}=\frac{\bar{P_s}}{P_{ASE}(B_m)}\times\frac{B_m}{B_r}\;\;\;\Big{(}P_{ASE}(B_m)=Tr(O^T_m\widehat{PG}O_m)\sigma^2=Tr(\widehat{PG}O_mO^T_m)\sigma^2\Big{)}.
\label{OSNR_NL_1}
\end{equation} 
Here $O_m$ is the low-pass transfer function of the  bandpass filter (bandwidth $B_m$).   
In Eq. (\ref{OSNR_NL_1}), $\sigma^2$ and $\widehat{PG}$ are given by Eq. (\ref{auto_noise_f3}) and Eq. (\ref{PG_K}), respectively. 

In the case of traditional OSA-based out-of-band OSNR monitoring, the ASE power can be interpolated using    
\begin{equation}
P_{ASE}(B_m, \pm \Delta \lambda)=
\frac{Tr\Big{[}\widehat{PG}\Big{(}O_m(-\Delta \lambda)O^T_m(-\Delta \lambda)+O_m(+\Delta \lambda)O^T_m(+\Delta \lambda)\Big{)}\Big{]}\sigma^2}{2}, 
\label{ASE_Delta}
\end{equation}
where $O_m(\pm \Delta \lambda)$ is the filter function centered at $\pm \Delta \lambda$.
When  $\Delta \lambda=0$, Eq. (\ref{ASE_Delta}) returns to the ASE power in  Eq. (\ref{OSNR_NL_1}), where
$P_{ASE}(B_m)\equiv P_{ASE}(B_m,0)$. 

\begin{figure}
\vskip -0mm
\begin{center}
    \begin{tabular}{cc}
     \resizebox{100mm}{!}{\includegraphics{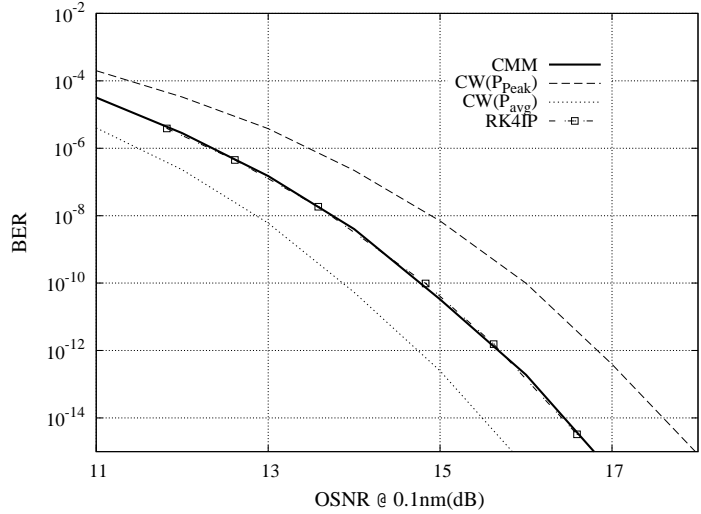}} \\ 
     \end{tabular}
    \vskip  -2mm
\caption{BER versus received OSNR for a 20Gb/s 20-span RZ-DPSK system with
         $\bar{\Phi}_{NL}=0.2 \pi$. Solid: obtained using CMM of Ref. \cite{Holzlohner02}.
         Dashed (dotted): improved  CW approach of Ref. \cite{Serena06} with
         CW power being 
         peak power (average power), respectively.
         Dash-dotted: RK4IP approach.
         All curves, except the dash-dotted, are obtained from Fig. 8 of Ref. \cite{Serena06}.
            }
    \label{fig_compare_Paolo_2}
\end{center}
\end{figure}

\section{Applications to DPSK systems}
\label{N_verification}
To show that the new approach to get the noise propagator is numerically applicable,
we will compare our RK4IP results with the CMM results given by Ref. \cite{Serena06} and with the experimental data
given by Ref. \cite{Coelho09}. Both consider systems with $R_b=$20 Gb/s, using RZ-50$\%$ DPSK modulation.
In the receiver, the optical filter is Gaussian type, while the electric filter is the fifth-order Bessel type.
In the following calculations, we set $T_0=NT_b$ by changing $\mu$ in Eq. (\ref{LMT_0}).
This means, given the noise propagator matrix,  the computational cost for BER is much higher than that in the linear case.
For BER calculations, the length of the de Bruijn sequence is $N=2^5$ \cite{Forestieri00}.
Based on the relation between the RK4IP step and the fiber dispersion length ($L_D$) or nonlinear length ($L_N$) discussed in Ref. \cite{zcb10}
as well as the detailed fiber parameters  in the following discussion,
we let the RK4IP step for the transmission fiber ($h_{tr}$)  and that for the DCF fiber ($h_{DCF}$) be related with
$h_{tr}:h_{DCF}=(5\sim6):1$. The value of $h_{tr}$ and the required computational time will be detailed below.

\subsection{Comparison with CMM results}
\label{Comp_CMM}

We consider the
20-span DPSK system discussed in Ref. \cite{Serena06}, where BERs using the CMM and the (improved) CW approaches 
were plotted against the received OSNR  in the Fig. 8 of Ref. \cite{Serena06}. 
In fact this system is basically the same as the one shown in 
Fig. \ref{fig_link} in Appendix A, 
provided that one removes the pre- and postcompensating fibers and their amplifiers 
in the Fig. 2 of Ref. \cite{Serena06} and removes   
the first EDFA and $N_{in}$ in our Fig. \ref{fig_link}. Thus the first term of $\widehat{PG}$ [in Eq. (\ref{PG_K})]
needs to be ignored.
Like Refs. \cite{Serena06,Serena05}, where OSNR was calculated in the absence of PG,   we obtain OSNR from Eq. (\ref{OSNR_L}) with 
$N_{in}=0$ and $(K+1)$ being replaced by $K$.
According to Ref. \cite{Serena06},
we change OSNR by changing the $n_{sp}$ in Eq. (\ref{PSD_ASE}).
As plotted in  Fig. \ref {fig_link}, each span contains a transmission fiber followed by
a dispersion-compensating fiber (DCF). The transmission fiber is $l=$100 km long with its CD parameter $D_{tx}=8$ ps/nm/km.
Each span is fully compensated. The nonlinear phase accumulated in the fiber, 
defined as $\bar {\Phi}_{NL}=\int_0^{z} \gamma (\xi) P_{in} e^{-\alpha(\xi)\xi}d\xi$ with $P_{in}$ being the time averaged signal power (at the input of the fiber), is  
$0.2\pi$.
The bandwidth of the optical (electrical) filter in the receiver  is $B_o=1.8R_b$ ($B_e=0.65R_b$), respectively.

To let our results be reproducible, we provide, as detailed  as possible, other related parameters below. 
The DCF in each span is 8 km long with $D_{DCF}=-100$ ps/nm/km.  
Transmission fiber and DCF are assumed to have same fiber loss ($\alpha=$0.2 dB/km)
and same nonlinear coefficient ( $\gamma=$2.0 /W/km).
The EDFA in each span  is used to compensate the total loss in the fiber of $L=(100+8)$ km.  
Therefore the signal  power at the input of each span ($P_{in}$)  keeps constant. 
Ignoring the nonlinear phase contribution of DCF \cite{Bononi08}, we set $P_{in}$=0.7307 mW, obtained from  
$\bar {\Phi}_{NL}=K \gamma P_{in}(1-e^{-\alpha l})/\alpha=0.2\pi$ with $K=20$ and $l=100$ km. 
The OSNR is obtained using Eq. (\ref{OSNR_L}) with $B_m/B_r=1.35$.


As shown in Fig. \ref{fig_compare_Paolo_2}, the curve using the proposed RK4IP approach (dash-dotted)
agrees very well with the
CMM curve (solid) given by Ref. \cite{Serena06}.
In Fig. \ref{fig_compare_Paolo_2}, the curves using improved CW approach \cite{Serena06} with CW power being transmitted peak power (dashed)
and average power (dotted) are plotted for comparison.

For each calculated point in Fig. \ref{fig_compare_Paolo_2}, the CPU time for the noise propagator calculation is $\sim$1.8 hr 
with  RK4IP step for transmission fiber ($20\times$100 km long) being $h_{tr}=$3.5 km. The CPU time for each BER calculation is $\sim$ 0.5 hr.
In fact,  $h_{tr}$ ranged  within $0.3 \sim 5$ km yields almost the same curve.
 
\begin{figure}
\vskip -0mm
\begin{center}
    \begin{tabular}{cc}
  \resizebox{120mm}{!}{\includegraphics{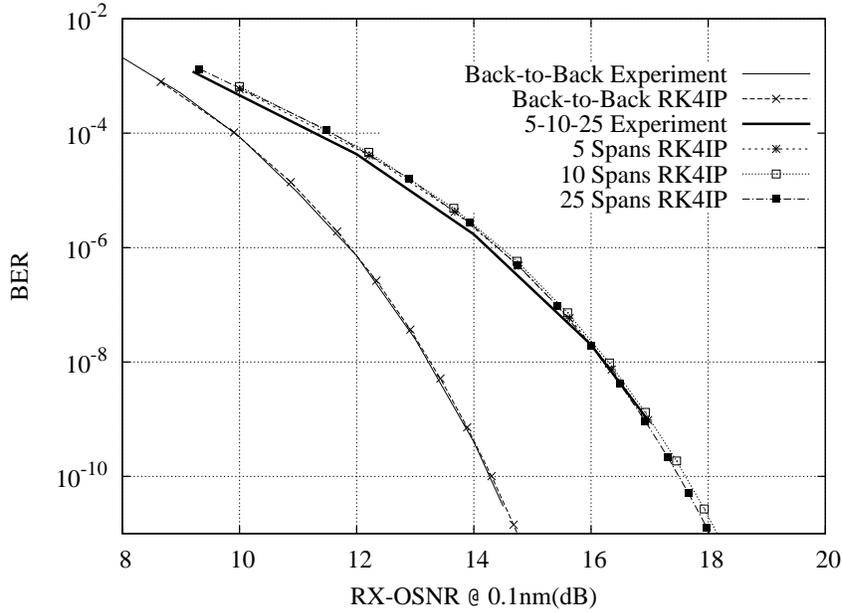}} \\
     \end{tabular}
    \vskip  -2mm
\caption{BER versus received OSNR for the multi-span RZ-50$\%$ DPSK systems with
         $\bar{\Phi}_{NL}=0.9$. 
         Each span consists of a SMF fiber (42 km long) and a DCF file (7 km long).
         Other fiber parameters were detailed in Table 1 of Ref. \cite{Coelho09}.
         According to Fig. 7 (b) in Ref. \cite{Coelho09}, where the experimental curves for 5-, 10-, and 25-span systems
         were almost the same, here we replot these three curves 
         using a thick solid curve.
            }
    \label{fig_compare_Coelho_1}
\end{center}
\end{figure}
\begin{figure}
\vskip -0mm
\begin{center}
    \begin{tabular}{cc}
  \resizebox{120mm}{!}{\includegraphics{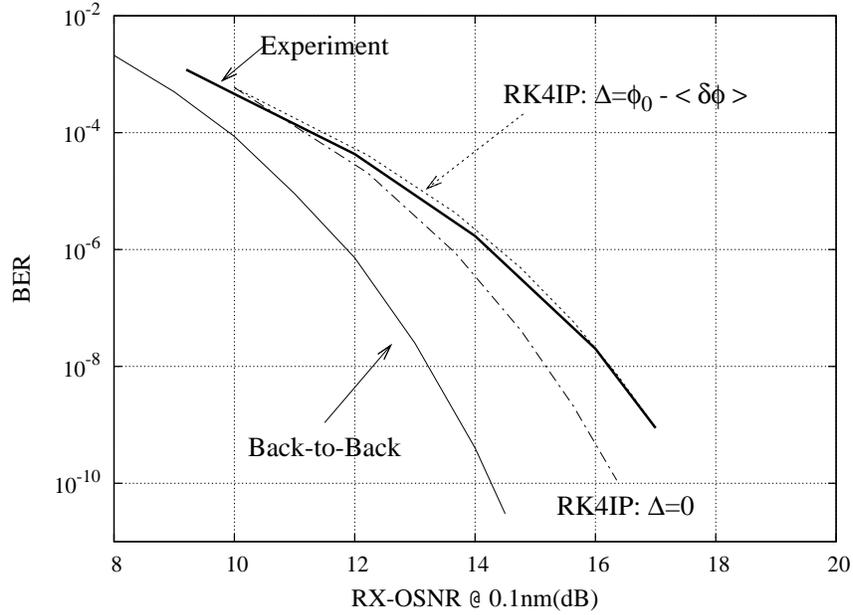}} \\
     \end{tabular}
    \vskip  -2mm
\caption{BER vs RX-OSNR for the 5-span RZ-50$\%$ DPSK system with 
         $\bar{\Phi}_{NL}=0.9$. 
         Other parameters are same as those used in Fig. \ref{fig_compare_Coelho_1}.
         Solid: experimental results. 
         Dotted (Dash-dotted): numerical calculation using RK4IP with (without) phase shift $\Delta$.
            }
    \label{fig_compare_Coelho_2}
\end{center}
\end{figure}

\subsection{Comparison with experimental data}
\label{Comp_exp}

The optical system discussed in Ref. \cite{Coelho09} can be modelled  by Fig. \ref{fig_link}, except that each EDFA
should be replaced by an EDFA followed by an optical filter (Gaussian) $O_{lk}$ with bandwidth $B_{lk}=$5 nm. 
Also, the input noise  $N_{in}$ needs to be filtered by an optical filter $O_{in}$ (Gaussian,  $B_{in}=$3 nm).
As a result, in Eq. (\ref {PG_K}), the noise propagator $p_n\Big{(}(k+1)L,kL\Big{)}$ ($k=0,\cdots,K-1$) should be 
replaced with $O_{lk}p_n\Big{(}(k+1)L,kL\Big{)}$, $P_n(K,K)=1$ with $P_n(K,K)=O_{lk}$, and 
$G\sigma^2_{in}P_n(K,0)P^T_n(K,0)$ with  $G\sigma^2_{in}P_n(K,0)O_{in}O^T_{in}P^T_n(K,0)$.
Since we only consider the curves  
plotted in Fig. 7 (b) of Ref. \cite{Coelho09},
each fiber ($L$ km long) in Fig. \ref{fig_link} contains a SMF (42 km long) followed by a DCF (7 km long).
In our calculation, all the related fiber parameters are same as those given in Table 1 of Ref. \cite{Coelho09}.
In the receiver, the bandwidth of the optical filter is $1.87R_b$, while the bandwidth of the electrical filter
is $0.75R_b$.

We first consider the back-to-back case. 
Similar to  Ref. \cite{Coelho09},  we modify the 20Gb/s RZ-50$\%$ signal at the transmitter  by 
comparing its calculated spectrum \cite{Ip06} and its measured spectrum\cite{Coelho09}.
Due to that the input noise $N_{in}$ is filtered by $O_{in}$, the OSNR is calculated using Eq. (\ref{OSNR_NL_1}) 
with $B_m/B_r=0.95$, yielding the back-to-back RK4IP curve shown in
Fig. \ref{fig_compare_Coelho_1}.  

For the 5-, 10-, 25-span systems, their accumulated nonlinear phases are   calculated according to 
Eq. (48) of Ref. \cite{Coelho09}. 
Because of the spectral modification of the input signal, the optical power at the input  of each span
$P_{in}=P_{SMF}$ is smaller than $E_b/T_b$, where $E_b$ is the energy per bit before the spectral modification.
For example, to get nonlinear phase $\bar{\Phi}_N=0.9$ for the 25-span system, 
the fiber input power $P_{in}=P_{SMF}$ should be 1.516 mW, which means $E_b/T_b=$0.127 mW or $G(E_b/T_b)=2.316$ mW ($G=18.197$).
Different from the DPSK receiver shown in Fig.\ref{fig_link}, where the delay is  $T_b=1/R_b=50$ ps, 
the delay in the receiver of Ref. \cite{Coelho09} was 
$T'_b$=(24.84 GHz)$^{-1}$=40.26 ps.
Thus, the DPSK phase factors given in Eq. (\ref{R_DPSK}) should be modified as
\begin{equation}
D^{ss}_{l'l}=\frac{e^{j\frac{2\pi l'}{N'}}\!+\!e^{-j\frac{2\pi l}{N'}}}{2},\;\;
D^{nn}_{m'm}=\frac{e^{j\frac{2\pi m'T'_b}{T_0}}\!+\!e^{-j\frac{2\pi mT'_b}{T_0}}}{2},\;
D^{ns}_{ml}=\frac{e^{j\frac{2\pi mT'_b}{T_0}-j\Delta}+e^{-j\frac{2\pi l}{N'}+j\Delta}}{2},
\label{R_DPSK_2}
\end{equation}
with $N'=N(T_b/T'_b)$, $T'_b=T_b+\Delta T_b$.
In Eq. (\ref{R_DPSK_2}), $\Delta$ 
is introduced as
\begin{equation}
\Delta=\phi_0-<\delta \phi>, 
\label{GM_shift}
\end{equation}
where $<\delta \phi>$, given by Eq. (\ref{d_phi_1}) in Appendix C, 
is the nonlinear phase difference  
between noise and noise-free signal.
As shown in Fig.  \ref{fig_compare_Coelho_1}, all RK4IP curves ($\phi_0=0.31$) agree very well with the experiment results.
The ASE power is calculated using Eq. (\ref{ASE_Delta}) with $\Delta \lambda=2B_m$. 
In Eq. (\ref {GM_shift}),  $\phi_0$ is  a calibration constant that basically shifts the RK4IP curves in the OSNR direction, 
while 
$<\delta \phi>$
determines the slope of the RK4IP curves. 
To show this, we plot in Fig. \ref{fig_compare_Coelho_2} the RK4IP results for the 25-span system with 
$\Delta=0.31-<\delta \phi>$ 
and $\Delta=0$.
Also, we consider the RK4IP curves using Eq. (\ref{OSNR_NL_1}) to calculate ASE power. 
Our results for the 5-span, 10-span, and 25-span systems confirm that there is  almost no difference between the 
curve using Eq. (\ref{ASE_Delta}) with 
$\phi_0=0.31$
and the curve using Eq. (\ref{OSNR_NL_1}) with $\phi_0=0.57$.

In Eq. (\ref{GM_shift}), the calibration constant $\phi_0$
can be temporally considered as a fitting parameter. As mentioned above, it basically affects the BER vs OSNR curve in the OSNR direction and
is related with the detailed OSNR monitoring technique used in Ref. \cite{Coelho09}.
As there is few information about its OSNR measurement, evaluation of $\phi_0$ is expected to be discussed  elsewhere.

In the above calculation, our CPU time for the noise propagator calculation is $\sim$0.8 hr
with  RK4IP step for transmission fibers ($25\times$40 km long) being $h_{tr}=$6.0 km. The CPU time for each BER calculation is $\sim$ 0.5 hr.
The step size $h_{tr}$  within $0.3 \sim 10$ km will result in almost the same curve.

\section{Summary}

For  linear OFC systems,  MGF method is useful for one to evaluate
the noise impacts on BERs. This is true  not only
because it is computationally efficient for the cases with low BER (e.g.,  $< 10^{-9}$)
but also it can provide reliable information  for the cases using coherent detection,
which is now widely used in modern OFC systems.
It is now well recognized that
traditional Gaussian fitting Q-factor approximation is accurate for OOK detection, while MGF method is accurate for various linear OFC systems.

To extend 
the MGF method 
to a nonlinear OFC system,
one needs to make sure 
its noise propagator varies within linear regime.
This means  the noise-noise interaction needs to be neglected and the noise field should  follow  LNE.

In CMM of Refs. \cite{Holzlohner02,Holzlohner03,Holzlohner03b},
the noise propagator was obtained from NLSE.
It 
may contain the
nonlinearity-induced  jitter, which is beyond the linear regime and should
be removed.
In this work we simplify the CMM by directly solving the accurate LNE [Eq. (\ref{NS_EDFA_IP})] with RK4IP.
Like the CW approximations (cf. Refs. \cite{Kikuchi93}-\cite{Secondini09} and many others) as well as the approach of
Ref. \cite {Demir04},
where the noise propagation information obtained from the related linearized noise equation
is automatically free from nonlinearity-induced jitter,
the noise propagator obtained from accurate LNE 
also varies within linear regime.

To numerically verify this new approach, we consider a 20-span RZ-DPSK system discussed in Ref. \cite{Serena06}.
The BERs obtained using this new RK4IP agree well with those using CMM in Ref. \cite{Serena06}.
Taking account the  phase difference between noise and noise-free signal
leads to quantitative matching between numerical
evaluation and experimental result \cite{Coelho09}.

\vskip 2mm
\noindent
{\bf Appendix A: Accurate LNE in the EDFA-based systems}
\vskip 2mm
\noindent

\begin{figure}
\vskip -0mm
\begin{center}
    \begin{tabular}{cc}
     \resizebox{120mm}{!}{\includegraphics{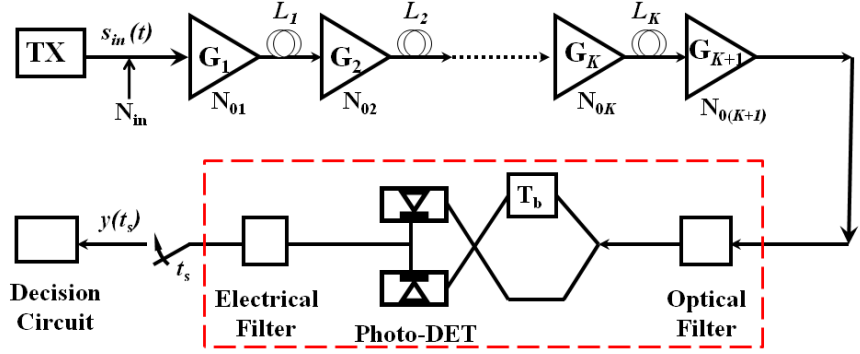}} \\ 
     \end{tabular}
    \vskip  -2mm
\caption{Low-pass equivalent optical model. 
         The receiver consists of optical and electrical filters and  DPSK balance detection.
         Our calculations, except those in Sec. \ref{Comp_CMM}, are based on the assumptions
         that $N_{0k}=N_0$ and $G_k=G$ [$k=1,2,...(K+1)$].
         Also, the fiber in each span is assumed to have same length ($L$ km) and same loss ($\alpha$ dB/km).
         $\mathrm{N}_\mathrm{in}$ is the ASE noise added at the transmitter. 
         Changing $\mathrm{N}_\mathrm{in}$ will change the OSNR at the receiver.
         In a typical balanced DPSK receiver, the 
         delay in one branch of the interferometer is $T_b=1/R_b$. In this work, $R_b=20$ Gb/s.
            }
    \label{fig_link}
\end{center}
\end{figure}


The optical field $u(z,t)$ in a fiber satisfies
\begin{eqnarray}
\frac{\partial u}{\partial z}
=j\frac{\beta_{\omega\omega}}{2}\frac{\partial^2 u}{\partial t^2}
+\frac{\beta_{\omega\omega\omega}}{6}\frac{\partial^3 u}{\partial t^3}
-j\gamma |u|^2 u-\frac{\alpha}{2}u,
\label{NLSE}
\end{eqnarray}
where $\alpha$ is the fiber loss and
$\beta_{\omega\omega}=\partial^2 \beta/\partial \omega^2$ relates  to the
CD parameter
$D(\lambda)$ (ps/nm/km) with 
$\beta_{\omega\omega}$=-$\frac{\lambda^2D(\lambda)}{2\pi c}$ (c=3$\times$10$^{8}$ m/s).
The slope parameter
$\beta_{\omega\omega\omega}=(\frac{\lambda}{2\pi c})^2[2\lambda D(\lambda)+\lambda^2D'(\lambda)]$
[$D'(\lambda)=\frac{dD(\lambda)}{d\lambda}$ (ps/nm$^2\cdot$km)] can be neglected if bit rate $R_b$
satisfies
$R_b>|\beta_{\omega\omega}/\beta_{\omega\omega\omega}|$ \cite{Forestieri00}.

Introducing the transformation  $u(z,t)=v(z,t)e^{-\alpha z/2}$, Eq. (\ref{NLSE}) can be reduced as
\begin{eqnarray}
\frac{\partial v}{\partial z}
=j\frac{\beta_{\omega\omega}}{2}\frac{\partial^2 v}{\partial t^2}
+\frac{\beta_{\omega\omega\omega}}{6}\frac{\partial^3 v}{\partial t^3}
-je^{-\alpha z} \gamma |v|^2 v.
\label{NLSE_2}
\end{eqnarray}
In  a $K$-span system amplified by $(K+1)$ EDFAs (cf. Fig. \ref{fig_link}), Eq. (\ref{NLSE_2}) can be modified as
\begin{eqnarray}
\frac{\partial v}{\partial z}
=j\frac{\beta_{\omega\omega}}{2}\frac{\partial^2 v}{\partial t^2}
+\frac{\beta_{\omega\omega\omega}}{6}\frac{\partial^3 v}{\partial t^3}
-je^{-\alpha z}\gamma |v|^2 v -jw(z,t),
\label{NLSE_ASE}
\end{eqnarray}
where  $w(z,t)$ is the ASE forcing modeled as the complex AWGN with correlation 
\begin{eqnarray}
r(z,z',t,t')=E\{w(z,t)w^*(z',t')\}=\delta(t-t')\delta(z-z')
\sum_{k=1}^{K+1} N_{0k}\delta(z-(k-1)L).
\label{correlat_t}
\end{eqnarray}
In Eq. (\ref{correlat_t}), the fiber length in each span is assumed to be $L$ (km) long,
According to Wiener-Kinchine theorem \cite{Papoulis91},
$N_{0k}$ in Eq. (\ref{correlat_t}) is the ASE PSD (in one polarization direction)
at the output of the $k$th EDFA.
Suppose each EDFA has the same gain $G$ and spontaneous-emission parameter
$n_{sp}$,
we have \cite{Forestieri00}
\begin{equation}
N_{0k}=N_0=n_{sp}(G-1)\hbar\omega \;\;\;\;\;\;\Big{(}k=1,2,...K,(K+1)\Big{)}
\label{PSD_ASE}
\end{equation}

Decomposing 
optical field $v(z,t)$ in the fiber into noise-free  field $v_0(z,t)$ and its perturbation
$\delta v(z,t)$ [i.e., $v(z,t)=v_0(z,t)+\delta v(z,t)$]
and assuming that $|v_0|>> |\delta v|$ (so that
the nonlinear terms of $\delta v$ can be neglected),
Eq. (\ref {NLSE_ASE}) can be decomposed as \cite{Holzlohner02}
\begin{eqnarray}
\frac{\partial v_0}{\partial z}
=j\frac{\beta_{\omega\omega}}{2}\frac{\partial^2 v_0}{\partial t^2}
+\frac{\beta_{\omega\omega\omega}}{6}\frac{\partial^3 v_0}{\partial t^3}
-je^{-\alpha z}\gamma  |v_0|^2 v_0
\label{NLSE0}
\end{eqnarray}
\begin{eqnarray}
\frac{\partial \delta v}{\partial z}
=j\frac{\beta_{\omega\omega}}{2}\frac{\partial^2 \delta v}{\partial t^2}
+\frac{\beta_{\omega\omega\omega}}{6}\frac{\partial^3 \delta v}{\partial t^3}
-j2e^{-\alpha z}\gamma |v_0|^2 \delta v -je^{-\alpha z}\gamma v_0^2 \delta v^*-jw(z,t).
\label{NLSE_ASE2}
\end{eqnarray}
Noise equation Eq. (\ref {NLSE_ASE2}) differs from common equations in  
that the real and imaginary parts of the
complex noise field need to be treated separately \cite{Holzlohner02}.

Denoting
$a_l=a(\omega_l)=\int \delta v e^{-j\omega_l t}dt$ 
and the circulant matrices $[M_{\nu}]_{lm}=\nu_{l-m}$, $[M_{\mu}]_{lm}=\mu_{l+m}$ with
\begin{equation}
\nu_l=\nu(\omega_l)=e^{-\alpha z} \int |v_0|^2 e^{-j\omega_l t}dt,
\;\;\;\;
\mu_l=\mu(\omega_l)=e^{-\alpha z} \int v_0^2 e^{-j\omega_l t}dt,
\label{hermit}
\end{equation}
in frequency domain,
Eq. (\ref {NLSE_ASE2}) has the form
\begin{eqnarray}
\frac{d a_l}{d z}
=-j\frac{\beta_{\omega\omega}}{2}\omega_l^2 a_l
-j\frac{\beta_{\omega\omega\omega}}{6}\omega_l^3 a_l
-j2\gamma(z) [M_{\nu}]_{lm}a_m
-j\gamma(z) [M_{\mu}]_{lm}a_m^*
-jW_l,
\label{NS_g2}
\end{eqnarray}
where $W_l=W(z,\omega_l)$ is the Fourier component of the forcing term $w(z,t)$ in Eq. (\ref{NLSE_ASE}). 
As indicated in Ref. \cite{Holzlohner02},
since $|v_0|^2$ in (\ref {hermit}) is real, $\nu_{l}=\nu_{-l}^*$.
So $M_{\nu}$ in (\ref{NS_g2}) is Hermitian, or,
its real  part $M_{\nu}^R$   is
symmetric,
while its imaginary part $M_{\nu}^I$ is anti-symmetric.
Also, as $[M_{\mu}]_{km}=\mu_{k+m}$,
both the real ($M_{\mu}^{R}$) and imaginary parts ($M_{\mu}^{I}$) of $M_{\mu}$ 
are symmetric.

The matrix form of Eq. (\ref{NS_g2}) is
\begin{eqnarray}
&&\frac{d a}{d z}=\bar{L}a+\nu a+\mu a^*-jW
\label{NS_g_mtrx}
\\
&&\bar{L}=jL_{CD},\;\;\;\nu=-2j\gamma(z)(M_{\nu}^R+jM_{\nu}^I),\;\;\mu=-j\gamma(z)(M_{\mu}^R+jM_{\mu}^I)
\label{M_decomp}
\end{eqnarray}
Introducing $\tilde {a}=(a_R, a_I)^T$ (for $a=a_R+ja_I$) and $\tilde{W}=(W_I, -W_R)^T$ (for $-jW=W_I-jW_R$),
Eq. (\ref {NS_g_mtrx}) is equivalent to
\begin{eqnarray}
\!\!\!&\!\!\!&\!\!\! \frac{d \tilde{a}}{d z}
=(\hat {L}+ \hat {\nu}+\hat {\mu})\tilde {a} +\tilde{W}
\label{NS_tilde}
\\
\!\!\!&\!\!\!&\!\!\!
\hat {L}=\left(\begin{array}{cc} 0 & -L_{CD}\\ L_{CD} & 0 \end{array}\right),
\;\;
\hat {\nu}=\left(\begin{array}{cc} \nu_{AA} & -\nu_{SS}\\ \nu_{SS} & \nu_{AA} \end{array}\right),
\;\;
\hat {\mu}=\left(\begin{array}{cc} \mu_R & \mu_I\\ \mu_I & -\mu_R \end{array}\right),
\label{hat_L_nu_mu}
\end{eqnarray}
with
$(L_{CD})_{ij}=-[\frac{\beta_{\omega\omega}}{2}\omega^2+\frac{\beta_{\omega\omega\omega}}{6}\omega^3]\delta_{ij}$,
$\nu_{AA}=2\gamma M_\nu^I$, $\nu_{SS}=-2\gamma M_\nu^R$,
$\mu_R=\gamma M_\mu^I$, and $\mu_I=-\gamma M_\mu^R$.
According to the discussion given below Eq. (\ref{NS_g2}), $\hat{\nu}$ ($\hat{\mu}$) in Eq. (\ref {hat_L_nu_mu}) is
antisymmetric (symmetric), respectively.
Calculation of the Kerr term 
$(\hat {\nu}+\hat {\mu})$ according to
Eq. (\ref{hat_L_nu_mu}) has the computational complexity much less than $O(N_W^3)$, where
$N_W$ is the number of Fourier components used for signal representation.
In fact, 
the computational cost of this way 
is basically determined by the FFTs in Eq. (\ref{hermit}), 
which has the computational complexity of
$O(N_W log N_W)$.

In frequency domain, ASE correlation relation (\ref{correlat_t})  has its matrix form
( $\tilde{W}(\omega_l) \rightarrow \tilde{W}_l\sqrt{\Delta f} $ )
\begin{eqnarray}
E\{\tilde{W}_l(z)\tilde{W}_{l'}^*(z')\}=\delta_{z,z'}\delta_{l,l'}\sum_{k=0}^{K}\frac{N_{0}}{2T_0}\delta_{z,kL}\;\;\;\;(l=1,\cdots,4M_n+2),
\label{auto_noise_f_W}
\end{eqnarray}
where Eq. (\ref{PSD_ASE}) has been used.
In Eq. (\ref{auto_noise_f_W}), 
$T_0=1/\Delta f$ 
and $M_n$ are  given by Eq. (\ref{LMT_0}). 
Eq. (\ref{auto_noise_f_W}) means that 
Eq. (\ref{NS_tilde}) can be equivalently replaced by 
\begin{equation}
\frac{d \tilde{a}}{d z}
=(\hat{L}+\hat{N})\tilde{a}; \;\;\;\;(\hat{N}=\hat{\nu}+\hat{\mu})
\label{NS_EDFA}
\end{equation}
with boundary condition
\cite{Forestieri04}
\begin{eqnarray}
E\{\tilde{W}(z_f)\tilde{W}^{*}(z_f)\}=\frac{N_{0}}{2T_0}I
\equiv \sigma^{2}I 
\label{auto_noise_f3}
\end{eqnarray}
with $I$ being a $(4M_n+2)\times(4M_n+2)$ unit matrix and $\sigma^2$ being the variance of the real or imaginary part of input ASE.

\vskip 2mm
\noindent
{\bf Appendix B: Filtered photoelectric current expressed using KLSE}
\vskip 2mm
\noindent
Given a linear optical system,
based on the discussions in Ref. \cite{Forestieri00} and the notations introduced in Ref. \cite{zcb07b},
the filtered photoelectric current $I(t)$ can be expressed in the form of
$I(t)=[\langle s^o(t+T_b)+n^o(t+T_b)|s^o(t)+n^o(t)\rangle + c.c.]/2$.
Here $s^o(t)$ ($n^o(t)$) represents the signal (noise) field at the input of the optical filter.
Dirac bra $\langle x|$ is the conjugate transpose (or Hermitian transpose) of Dirac ket $|x\rangle$ [$x=s^o(t), n^o(t), s^o(t)+n^o(t), etc.$].
The Dirac ket differs from usual complex vector in that the $i$th element of the latter is just the $i$th Fourier coefficient of
the (signal or noise) field,
while the $i$th element of the former is the product of the $i$th Fourier coefficient and its base function
(cf. Eq. (17) in Ref. \cite{zcb07b}).
According to Refs. \cite{Forestieri00,zcb07b}, 
the filtered current in a linear system can be formally expressed  as
$I(t)=y_{ss}+y_{nn}+y_{ns}$ with ($l=-L_s,\cdots L_s$; $m=-M_n\cdots M_n$)
\begin{eqnarray}
y_{ss}(t_{\mathrm{s}})\!\!\!&\!\!\!=\!\!\!&\!\!\![\langle s^o(t_{\mathrm{s}}+T_b)|R_{ss}|s^o(t_{\mathrm{s}})\rangle+c.c.]/2
=\langle s^o(t_{\mathrm{s}})|R_{ss}^D|s^o(t_{\mathrm{s}})\rangle
\nonumber \\                                                
y_{nn}(t_{\mathrm{s}})\!\!\!&\!\!\!=\!\!\!&\!\!\![\langle n^o(t_{\mathrm{s}}+T_b)|R_{nn}|n^o(t_{\mathrm{s}})\rangle+c.c.]/2
=\langle N^o|R_{nn}^D|N^o\rangle=\langle Z|\Lambda|Z\rangle
\nonumber \\                                                  
y_{ns}(t_{\mathrm{s}})\!\!\!&\!\!\!=\!\!\!&\!\!\!
[\langle n^o(t_{\mathrm{s}}+T_b)|R_{ns}|s^o(t_{\mathrm{s}})\rangle+\langle n^o(t_{\mathrm{s}})|R_{ns}|s^o(t_{\mathrm{s}}+T_b)\rangle+c.c.]/2
\nonumber \\
\!\!\!&\!\!\!=\!\!\!&\!\!\!
[\langle N_{in}|O^{\dagger}_{nn}R_{ns}^D|s^o(t_{\mathrm{s}})\rangle+c.c]=[\langle Z|b^D(t_{\mathrm{s}})\rangle +c.c.]
\label{y_D}
\end{eqnarray}
where
$\Lambda\equiv U^{\dagger}O^{\dagger}_{nn}R_{nn}^DO_{nn}U=diag\{\lambda_1,\cdots,\lambda_{2M_n+1}\}$,
$|b^D(t_{\mathrm{s}})\rangle=U^{\dagger}O^{\dagger}_{nn}R_{ns}^D|s^o(t_{\mathrm{s}})\rangle$, and
\begin{eqnarray}
\!\!\!\!\!&\!\!\!\!\!&\!\!\!\!\!
(R_{ss}^D)_{l'l}=(R_{ss})_{l'l}D^{ss}_{l'l},\;\;(R_{nn}^D)_{m'm}\!=\!(R_{nn})_{m'm}D^{nn}_{m'm},\;\;
(R_{ns}^D)_{ml}=(R_{ns})_{ml}D^{ns}_{ml}
\nonumber \\
\!\!\!\!\!&\!\!\!\!\!&\!\!\!\!\!
D^{ss}_{l'l}=\frac{e^{j\frac{2\pi l'}{N}}\!+\!e^{-j\frac{2\pi l}{N}}}{2},\;\;
D^{nn}_{m'm}=\frac{e^{j\frac{2\pi m'T_b}{T_0}}\!+\!e^{-j\frac{2\pi mT_b}{T_0}}}{2},\;
D^{ns}_{ml}=\frac{e^{j\frac{2\pi mT_b}{T_0}}+e^{-j\frac{2\pi l}{N}}}{2}.
\label{R_DPSK}
\end{eqnarray}
In Eqs. (\ref {y_D})-(\ref{R_DPSK}),
$|Z \rangle$ represents the decoupled Gaussian random variables with zero mean and real part and imaginary part variance of
$\sigma^2$. 
The effects of the optical and electrical filters in the receiver are
represented by matrices with their elements being
$(O_{nn})_{mm'}=\delta_{m,m'}H_o(\frac{m}{T_0})$, $(O_{ss})_{ll'}=\delta_{l,l'}H_o(\frac{l}{NT_b})$ and
$(R_{ss})_{l l'}= H_r(\frac{l'-l}{NT_b}), (R_{nn})_{m m'}= H_r(\frac{m'-m}{T_0}), (R_{ns})_{m l}=H_r(\frac{l}{NT_b}- \frac{m}{T_0})$.
Due to the optical and electrical filters, signal (noise) components outside $\pm L_s$ ($\pm M_n$) can be neglected.
Here \cite{Forestieri00}
\begin{equation}
L_s=\eta NT_b B_o,\;\; M_n=\eta B_oT_0,\;\; T_0=\mu(\frac{1}{B_o}+\frac{1}{B_r})\;.
\label{LMT_0}
\end{equation}

For a nonlinear optical system, 
to get the noise propagator from the accurate
LNE (\ref{NS_tilde}),
one needs to separate complex numbers into their real and imaginary parts.
Denoting the Re-Im form of a complex matrix $x$ as
$\tilde{x}= \left(\begin{array}{cc} Re\{x\} & -Im\{x\}\\ Im\{x\} & Re\{x\} \end{array}\right)$
[where $x$ can be any complex matrix 
in Eq. (\ref{y_D})]
and
introducing 
$|n^o\rangle=P_{n,eq}|a_0\rangle$ with $|a_0\rangle$ being the AWGN from EDFA,
\begin{eqnarray}
|\tilde{s}^o \rangle=\left[\begin{array}{c} Re\{ |s^o\rangle \}\\ Im\{ |s^o\rangle \} \end{array}\right],
\;\;
|\tilde{a}_0 \rangle=\left[\begin{array}{c} Re\{ |a_0\rangle \}\\ Im\{ |a_0\rangle \} \end{array}\right],
\;\;
|\tilde{Z} \rangle=\left[\begin{array}{c} Re\{ |Z\rangle \}\\ Im\{ |Z\rangle \} \end{array}\right]
,
\label{tilde_s_n}
\end{eqnarray}
it is easy to generalize the noise related currents in Eq. (\ref{y_D}) as
\cite{Holzlohner02, Serena06,Coelho09,Secondini09,Forestieri00}
\begin{eqnarray}
\!\!\!\!\!\!\!\!&\!\!\!\!\!\!\!\!\!&\!\!\!\!\!\!\!\!
y_{nn}(t_{\mathrm{s}})
\!\!=\!\!\langle \tilde{a}_{0}|\tilde{U}\tilde{\Lambda}\tilde{U}^T |\tilde{a}_{0}\rangle
\!\!\equiv\!\!\langle \tilde{Z}|\tilde{\Lambda}|\tilde{ Z}\rangle \;\;\;\;\;
(\tilde{\Lambda}\!\!=\!\!\tilde{U}^TP_{n,eq}^T\tilde{O}_{nn}^{T}\tilde{R}^D_{nn}\tilde{O}_{nn}P_{n,eq}\tilde{U}
\!\!=\!\!diag\{\tilde{\lambda}_1,\cdots, \tilde{\lambda}_{4M_n+2}\})
\nonumber \\
\!\!\!\!\!\!\!\!&\!\!\!\!\!\!\!\!\!&\!\!\!\!\!\!\!\!
y_{ns}(t_{\mathrm{s}})
\!\!=\!\!\langle\tilde{Z}|\tilde{U}^TP_{n,eq}^T \tilde{O}_{nn}^{T}\tilde{R}^D_{ns}\tilde{O}_{ss}|\tilde{s}^o(t_{\mathrm{s}})\rangle
\!\!=\!\!\langle\tilde{Z}|\tilde{B} |\tilde{s}^o(t_{\mathrm{s}})\rangle\equiv \langle\tilde{Z}|\tilde{b}(t_{\mathrm{s}})\rangle
\;\;\;(\tilde{B}\!\!=\!\!\tilde{U}^TP_{n,eq}^T\tilde{O}_{nn}^{T}\tilde{R}^D_{ns}\tilde{O}_{ss}),
\label{y_receiv2}
\end{eqnarray}
where $P_{n,eq}$ can be obtained from  Eqs. (\ref{H_RK4IP})-(\ref{p_noise_L}) and (\ref{PG_K})-(\ref{PG_K_eq}).

\vskip 2mm
\noindent
{\bf Appendix C: Nonlinearity induced phase difference between noise and  noise-free signal}
\vskip 2mm
\noindent
{\bf I: The phase difference caused by $\mathrm{N_{in}}$}

In this part, we assume that, in Fig. \ref{fig_link}, the external noise injected at the transmitter
( $\mathrm{N_{in}}$ ) is much larger than the ASE noise from the EDFAs, which is true for the experiments discussed
in Ref. \cite{Coelho09}.
Thus, one can only consider the phase difference caused
by $\mathrm{N_{in}}$ and ignores  the effect of $\mathrm{N}_{0k}$ ($k=1,\cdots,K$).

It is well known that, for the noise-free signal with its path average power being $\bar {P}$,
its nonlinear phase accumulated at the fiber output  is
$\bar \Phi_{NL}=\bar {P}\gamma KL$, where $L$ is the fiber length of each span, as denoted in Fig. \ref{fig_link}.
Due to the optical power fluctuation  $\delta \bar {P}$, 
the actual nonlinear phase  becomes
\begin{equation}
\phi_{NL}=(\bar {P}+\delta \bar {P}) \gamma KL.
\label{phi_N_in}
\end{equation}
Relative to the noise-free signal, the noise-induced phase change, $\phi_{NL}-\bar \Phi_{NL}$, varies randomly.
The average variance of such phase noise can be calculated as
\begin{equation}
<\delta \phi^2>=\equiv <\phi_{NL}^2-\bar \Phi_{NL}^2>\approx 2 \bar {P}<\delta \bar {P}>(\gamma KL)^2. 
\label{d_phi_2}
\end{equation}
With the approximation of negligible $N_{0k}$ ($k=1\cdots K$), $<\delta \bar {P}>$ in Eq. (\ref{d_phi_2})  is
 basically caused by
$\mathrm{N_{in}}$ in Fig. \ref{fig_link}.
For the experiments in Ref. \cite{Coelho09}, $\mathrm{N_{in}}$ is filtered with bandwidth of $B_{in}$=3nm.
Thus, we have
$ <\delta \bar {P}>=G\mathrm{N_{in}}B_{in}$.
Eq. (\ref {d_phi_2}) yields
\begin{equation}
<\delta \phi>\approx \sqrt{2} \bar {P}\gamma KL/\sqrt{ \bar {P}/<\delta \bar {P}>}
=\sqrt{2} \bar \Phi_{NL}/\sqrt{OSNR},
\label{d_phi_1}
\end{equation}
where  $OSNR\approx \bar {P}/(G\mathrm{N_{in}}B_{in})$ is the input OSNR.

Considering that $\delta \bar {P} \ge 0$, we have $\phi_{NL}>\bar \Phi_{NL}$. This means
$\phi_{NL}$ rotates faster than $\bar \Phi_{NL}$ and
there is a phase difference between the actual optical field and the noise-free field.
As part of the actual field, the noise field also has the same phase shift relative to the noise-free signal.
Note that this phase shift will not affect
signal-signal and noise-noise beatings. But it will affect the signal-noise beating.
In fact, when calculating the signal-noise beating, the noise and signal should be treated consistently. Or,  they
should be considered within the same
coordinate system.
In this work, $<\delta \phi>$ given by Eq. (\ref{d_phi_1}) is approximated as the average of  such phase shift.
Our numerical results  plotted in Figs. \ref{fig_compare_Coelho_1} and \ref{fig_compare_Coelho_2} confirm this approximation.

For the experiments in Ref. \cite{Coelho09}, we have $\bar \Phi_{NL}=0.9$.
In this case, Eq. (\ref{d_phi_1}) yields 
\begin{equation}
<\delta \phi>\approx 1/\sqrt{OSNR}\approx \pi/2-\mathrm{arctan}(\sqrt{OSNR}),
\label{d_phi_1d}
\end{equation}
where $\mathrm{arctan}(x)+\mathrm{arctan}(1/x)=\pi/2$ and
$\mathrm{arctan}(x)\approx x$ (for $x\rightarrow 0$) have been used.
Obviously, $<\delta \phi>$ in Eq. (\ref{d_phi_1d}) relates $\phi_{GM}$ in Eq.(23) of Ref. \cite{arXiv_zcb12} with  $<\delta \phi>+\phi_{GM}=\pi/2$.
Also, $\phi_0$ in Ref. \cite{arXiv_zcb12} now becomes
$\phi_0+\pi/2\rightarrow \phi_0$ in Eq. (\ref{GM_shift}),
while $\Delta_{GM}$ in Ref. \cite{arXiv_zcb12} is named as $\Delta$ in this work.

\vskip 2mm
\noindent
{\bf II: The phase difference caused by $\mathrm{N_{in}}$ and $\mathrm{N}_{0k}$ ($k=1,\cdots,K$) }

For the $K$-span system of Fig.  \ref{fig_link}  with $N_{in} \neq 0$ and $K$ not being large enough
and with the ASE from each EDFA being filtered by $O_{lk}$ 
and the ASE injected at the transmitter 
being filtered by $O_{in}$, 
Eq. (\ref{d_phi_2}) can be generalized as 
\begin{equation}
<\delta \phi^2>=2\Big[ GN_{in}B_{in}+N_0 B_{lk}\frac{K}{3}(1+\frac{1}{K})(1+\frac{1}{2K})\Big ] \bar \Phi_{NL}^2/\bar {P},
\label{d_phi_2c}
\end{equation}
where  $\sum_{k=1}^K k^2=\frac{K^3}{3}(1+\frac{1}{K})(1+\frac{1}{2K})$ has been used.
Obviously, in the case of $GN_{in}B_{in}>> KN_0B_{lk}$ \cite{Coelho09},
Eq. (\ref{d_phi_2c}) yields Eq. (\ref{d_phi_1}).

\vskip 0.10 in
\noindent
{\bf Acknowledgment}

\vskip 0.10 in

\noindent
The authors acknowledge the financial support from Canada Research Chair program.
The first author sincerely thanks Paolo Serena and  Leonardo D. Coelho for
providing their detailed calculations  in Refs. \cite{Serena06, Coelho09}.
The authors wish to thank the anonymous reviewers for their valuable comments and suggestions.


\end{document}